\documentclass[manuscript]{aastex}
\usepackage{emulateapj5,apjfonts}
\tighten

\slugcomment{The Astronomical Journal, in press}

\begin{document}

\title{A {\it Chandra} Study of the Circinus Galaxy Point-Source Population}

\author{
F.~E.~Bauer,\altaffilmark{1}
W.~N.~Brandt,\altaffilmark{1}
R.~M.~Sambruna,\altaffilmark{1,2}
G.~Chartas,\altaffilmark{1}
G.~P.~Garmire,\altaffilmark{1}
S.~Kaspi,\altaffilmark{1}~and~H.~Netzer\altaffilmark{3}}

\altaffiltext{1}{Department of Astronomy and Astrophysics, 525 Davey Lab, 
The Pennsylvania State University, University Park, PA 16802.}

\altaffiltext{2}{Department of Physics \& Astronomy and School of Computational
Sciences, George Mason University, 4400 University Dr. M/S 3F3, 
Fairfax, VA 22030-4444.}

\altaffiltext{3}{School of Physics and Astronomy, Raymond and Beverly Sackler
Faculty of Exact Sciences, Tel-Aviv University, Tel-Aviv 69978, Israel.}


\begin{abstract}
We have used the {\it Chandra X-ray Observatory} to resolve spatially
and spectrally the X-ray emission from the Circinus Galaxy. We report
here on the nature of the X-ray emission from the off-nuclear point
sources associated with the disk of Circinus, which make up $\approx$
34\% of the total 0.5--10~keV emission. We find that many of the
serendipitous X-ray sources are concentrated along the optical disk of
the galaxy, although few have optical counterparts within 1\arcsec\ of
their X-ray positions down to limiting magnitudes of $m_{\rm
V}=$~23--25. At the distance of Circinus ($\approx$ 3.8 Mpc), their
intrinsic 0.5--10~keV luminosities range from $\approx 2\times
10^{37}$~erg~s$^{-1}$ to $\approx 4\times 10^{39}$~erg~s$^{-1}$.
One-fourth of the sources are variable over the duration
of the 67~ks observation, and spectral fitting of these off-nuclear
sources shows a diverse range of spectral properties. The overall
characteristics of the point sources suggest that most are X-ray
binaries and/or ultra-luminous supernova remnants within Circinus.

We are able to analyze the two strongest off-nuclear sources in
greater detail and find both to have remarkable properties. The
average X-ray luminosities of the two sources are
$3.7\times10^{39}$~erg~s$^{-1}$ and $3.4\times10^{39}$~erg~s$^{-1}$.
The former displays large and periodic flux variations every 7.5~hr
and is well fit by a multicolor blackbody accretion-disk model with
$T_{\rm in}=1.35$~keV, properties consistent with an eclipsing $\ga$
50 $M_{\odot}$ black-hole binary. The latter appears to be a young
supernova remnant, as it coincides with a non-thermal radio
counterpart and an $\mathrm{H}\alpha$-detected \ion{H}{2} region. This
source exhibits both long-term ($\approx$~4~yr) X-ray variability and
a 6.67--6.97~keV iron emission-line blend with a 1.6~keV equivalent
width. These two objects further support the notion that
super-Eddington X-ray sources in nearby galaxies can be explained by a
mixture of intermediate-mass black holes in X-ray binaries and young
supernova remnants.

\end{abstract}

\keywords{
galaxies: active --- 
galaxies: individual (Circinus) --- 
X-rays: galaxies --
X-rays: binaries}

\section{INTRODUCTION}\label{intro}

The study of X-ray emission from star-forming galaxies has long lagged
behind studies in more traditional wavebands (e.g., ultraviolet,
optical and radio) because of sensitivity and resolution limitations.
These constraints largely restricted detailed investigations to
sources in the Local Group, where only small pockets of ongoing star
formation exist. The {\it Einstein X-ray Observatory} was the first to
resolve the X-ray source populations in nearby galaxies
\citep[e.g.,][]{Helfand1984, Fabbiano1989}, opening the study of
luminous X-ray binaries (XBs) and supernova remnants (SNRs) to sources
outside our own Galaxy (and the Magellanic Clouds) where there is
generally less ambiguity in the distance to the source and hence its
intrinsic luminosity. The brightest point sources in star-forming
galaxies were found to be significantly more luminous on average than
the brightest sources in the Milky Way \citep[e.g.,][]{Fabbiano1989}.
More sensitive X-ray observations of normal and star-forming galaxies
with {\it ROSAT} and {\it ASCA} have shown that variable, off-nuclear
point sources with super-Eddington X-ray luminosities (i.e., $L_{\rm X}\ga
2\times10^{38}$~erg~s$^{-1}$ for $M=1.4M_{\odot}$) are common but
generally not well understood \citep[e.g.,][]{Colbert1999,
Makishima2000, Roberts2000, Lira2000}.

The large improvements in spatial resolution and sensitivity recently
afforded by {\it Chandra}, however, have begun to change this picture.
{\it Chandra} not only offers detailed studies of the brightest point
sources in neighboring galaxies, but also studies of these sources
over a broad range of galaxy types and luminosity classes, providing
large samples of X-ray point sources that can be related by luminosity
and spatial distribution to other properties of these galaxies such as
gas mass, star formation rate and morphology (e.g., proximity to
spiral arms, \ion{H}{2} regions or globular clusters). Recent {\it
Chandra} observations of M31 \citep{Garcia2000}, M81
\citep{Tennant2001}, M82 \citep{Griffiths2000}, NGC~3256 (P. Lira et
al., in preparation), NGC~4038/39 \citep{Fabbiano2001}, and NGC~4647
\citep{Sarazin2000} are just a few of the diverse examples.

Here we perform a high-resolution X-ray study of the massive spiral
known as the Circinus Galaxy (hereafter Circinus). Because of its
proximity to the Galactic plane ($b=-3\fdg8$), Circinus lay hidden
until the 1970s. \citet{Freeman1977} found that Circinus lies at a
distance of 3.8 $\pm$ 0.6~Mpc\footnote{Assuming $H_{\rm
0}=75$~km~s$^{-1}$~Mpc$^{-1}$. At this distance, 1\arcsec\ corresponds
to 19~pc. } with an inclination of $\sim65^{\circ}$ to the
line-of-sight and appears to be located within a Galactic ``window''
with a visual absorption of $A_{\rm V} = 1.5 \pm 0.2$ and a neutral
hydrogen column density of $N_{\rm H} =(3.0 \pm 0.3)\times10^{21}$~cm$^{-2}$
[whereas neighboring regions typically have $A_{\rm V} = 3.0$ and
$N_{\rm H} =$ (5--10)$\times10^{21}$~cm$^{-2}$; Schlegel, Finkbeiner,
\& Davis 1998; Dickey \& Lockman 1990].\footnote{The Galactic
absorption column for Circinus was calculated from $A_{\rm V}$
assuming the average ratio $N_{\rm H}/A_{\rm
V}=1.7\times10^{21}$~cm$^{-2}$~mag$^{-1}$ as determined by
\citet{Diplas1994a} and \citet{Lockman1995}. We caution that the 
actual ratio can deviate from the average by as much as a factor of two
\citep{Burstein1978}.} Since its discovery, numerous lines of evidence
have been presented indicating that this galaxy hosts a Seyfert
nucleus: a prominent [\ion{O}{3}] ionization cone \citep{Marconi1994,
Wilson2000}, near-IR polarized bipolar scattering cones
\citep{Ruiz2000}, a coronal line region
\citep{Oliva1994, Maiolino2000}, polarized broad $\mathrm{H}\alpha$
\citep{Oliva1998, Alexander2000} and strong X-ray iron $\mathrm{K}\alpha$
lines \citep{Matt1996, Sambruna2001b}. Additionally, this galaxy
displays a moderate level of star formation, as indicated by
$\mathrm{H}\alpha$ imaging of a complex network of \ion{H}{2} regions
\citep[e.g., $\sim10^{7-8}$~yr~old star-formation rings at $\sim40$
and $\sim200$~pc from the nucleus;][]{Marconi1994, Wilson2000}. Past
X-ray studies have concentrated on the obscured nuclear source, with
the only spatially resolved pre-{\it Chandra} observation of Circinus
being a 4~ks {\it ROSAT} High-Resolution Imager (HRI) detection of the
nucleus and one off-nuclear point source
\citep{Guainazzi1999}.

In this paper, we study the properties of 16 X-ray point sources
coincident with Circinus which have fluxes $F_{\rm
0.5-10.0~keV}>7\times10^{-15}$~erg~cm$^{-2}$~s$^{-1}$. Details of the
observation and reduction procedure are outlined in
$\S$\ref{reduction}, and source identifications are discussed in
$\S$\ref{detection}. In $\S$\ref{timing} and $\S$\ref{spectra} we
discuss timing and spectroscopic analyses of the X-ray sources. We
comment on the luminosity function of the point sources in Circinus in 
$\S$\ref{lf}, and, finally, we discuss the implications of our findings
in $\S$\ref{discussion}. 

\section{Observations and Data Reduction}\label{reduction}

Circinus was observed with the High-Energy Transmission Grating
Spectrometer (HETGS; C. R. Canizares et al. in preparation) on 2000
June 6--7 with ACIS-S (G. P. Garmire et al., in preparation) in the
focal plane. The HETGS provided both a zeroth-order image of Circinus
on the ACIS S3 CCD and high-resolution dispersed spectra from two
transmission gratings, the Medium-Energy Grating (MEG) and the
High-Energy Grating (HEG), along the ACIS S0--S5 CCDs. The Circinus
nucleus was placed near the aim point on the S3 chip and was observed
using a subarray window of 600 rows to reduce pileup in the
zeroth-order image.\footnote{Pileup is the coincidence of two or more
photons in a pixel per ACIS frame that are counted by the instrument
as a single, higher energy event. The presence of pileup results in
photometric inaccuracy as well as in distortions to the HRMA+ACIS
point spread function and to ACIS spectra.} This particular
configuration provided a frame time of 2.1~s and reduced pileup in
the nucleus (i.e., the brightest X-ray source in the field) to
$\approx$ 3\% and in the off-nuclear sources to $\la$ 1\%
\citep[see][]{Sambruna2001a}.\footnote{The frame time is the  
fundamental unit of exposure for readout of the ACIS CCDs. The default
frame time is 3.2~s, but it can range from 0.2--10.0~s, depending on
the number of ACIS chips and subarray used.} Events were telemetered
in Faint mode, and the CCD temperature was $-120^{\circ}$~C. {\it
Chandra} is known to experience periods of especially high background
which are most pronounced on the S3 chip,\footnote{See
http://asc.harvard.edu/cal/Links/Acis/acis/Cal\_prods/bkgrnd/10\_20/bg201000.html.}
but such ``flares'' are much less frequent when the HETGS is
inserted.\footnote{See http://space.mit.edu/HETG/flight/status.html.}
We found that no background ``flares'' occurred during this
observation.

Analysis was performed on reprocessed {\it Chandra} data (2000
December reprocessing) primarily using the {\tt CIAO} V2.0 software
provided by the {\it Chandra} X-ray Center (CXC), but also with {\tt
FTOOLS} and custom software. We removed the $0\farcs5$ pixel
randomization, performed standard {\it ASCA} grade selection, and
excluded bad pixels and columns. The observation was split into two
continuous segments separated by 1,270~s because the HETGS was not
fully inserted during the first segment ($\sim0\fdg3$ from full
insertion). While the HETGS misalignment had a profound effect on the
gratings data, changes to the zeroth-order imaging and spectroscopy
should be negligible.\footnote{See
http://space.mit.edu/HETG/technotes/offset\_effect\_000619.txt.} The
total net exposure times for these two segments were 6,933~s and
60,153~s, respectively. The two datasets were merged (67,086~s) to
enhance source detection and timing analysis. Since the telescope did
not move between the two exposures, we also extracted spectra using
the full exposure. Event PI values and photon energies were determined
using the latest gain files appropriate for the observation. The X-ray
spectra were analyzed using {\tt XSPEC} \citep{Arnaud1996}. Unless
stated otherwise, spectral parameter errors are for 90\% confidence
assuming one parameter of interest.

The X-ray fluxes and absorption-corrected luminosities for all sources
with more than 90 counts were calculated from spectral fitting using
{\tt XSPEC}. For sources below this count limit, fluxes and
luminosities were estimated assuming an average power-law spectrum
with photon index $\Gamma=1.76$ and $N_{\rm
H}=1.5\times10^{22}$~cm$^{-2}$ as determined from the best-fit values
to sources above the 90 count limit. These flux and luminosity
estimates are assumed throughout the paper unless stated otherwise.
The scatter in the observed values of $\Gamma$ and $N_{\rm H}$
suggests that there is likely to be some systematic uncertainty in our
low-count flux and luminosity estimates. To assess this potential
uncertainty, we replaced the average $\Gamma$ and $N_{\rm H}$ values
with the entire range of values for sources with more than 90 counts,
rather than the average values. This led to deviations of 
no more than 36\% above and 24\% below the average flux and 54\%
above and 15\% below the average absorption-corrected luminosity. We
also note that the average value of $N_{\rm H}$ used above is much
larger than the Galactic absorption column
($3\times10^{21}$~cm$^{-2}$) and has a large dispersion
($8\times10^{21}$~cm$^{-2}$), implying that there is likely to be a
wide range in $N_{\rm H}$ values among individual sources.

\section{Source Detection and Distribution}\label{detection}

\begin{figure*}
\vspace{-0.1in}
\centerline{\hglue-0.3in{\includegraphics[height=4.8in,angle=0]{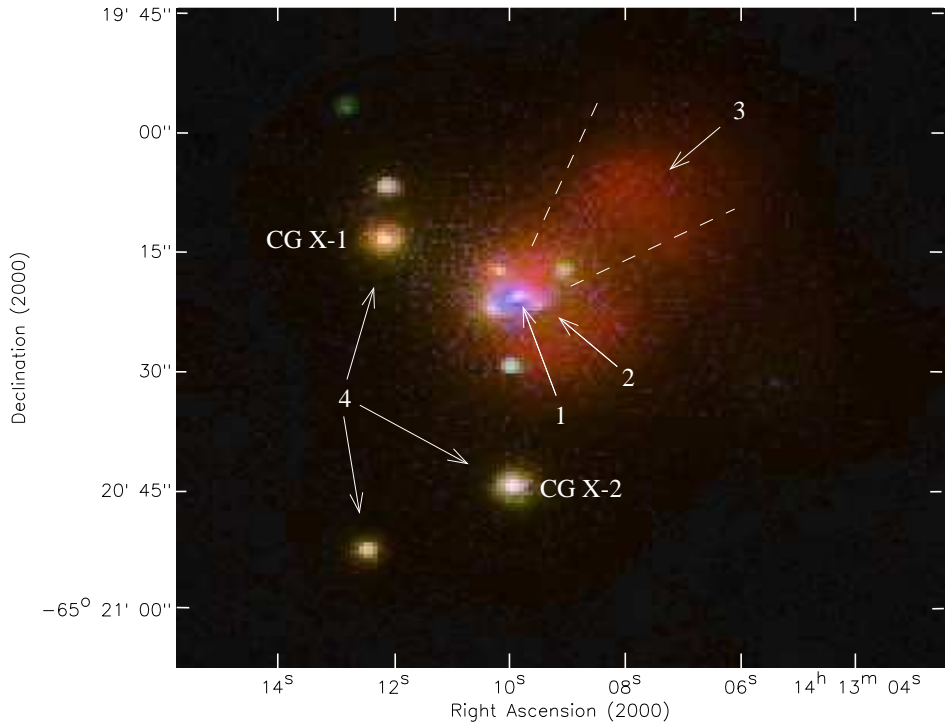}}}
\vspace{-0.7in}
\figcaption[fig1.eps]{An HETGS zeroth-order ``false color'' $80\arcsec
\times 80\arcsec$ image of Circinus with red, green, and blue representing
0.5--2.0~keV, 2.0--4.0~keV, and 4.0--10.0~keV emission, respectively.
Prior to combination, each color image was smoothed with an adaptive
kernel algorithm \citep{Ebeling2001} to enhance diffuse features. The
emission features depicted are (1) a central point source, (2) a
diffuse circumnuclear reflection halo, (3) a soft X-ray plume
coincident with the known ionization cone, and (4) several off-nuclear
point sources. Most of the off-nuclear sources appear to lie along the
projected major axis of the galaxy with a position angle
of about $-45^{\circ}$ East of North). The dashed lines represent the
opening angle of the [\ion{O}{3}] and $\mathrm{H}\alpha$ detected
ionization cone \citep{Marconi1994,Wilson2000}. The two brightest
off-nuclear X-ray sources, CG~X-1 and CG~X-2, are marked.
\label{fig:asmooth}}
\end{figure*}

Figure~\ref{fig:asmooth} shows an HETGS zeroth-order ``false color''
image of Circinus. The colors red, green, and blue represent the soft
(0.5--2.0~keV), medium (2.0--4.0~keV), and hard (4.0--10.0~keV) X-ray
emission, respectively. Prior to combination, each color image was
smoothed with an adaptive kernel algorithm \citep{Ebeling2001} which
permits the simultaneous viewing of compact and extended sources. The
smoothed images each have a signal-to-noise ratio of 2.5 per smoothing
beam and were corrected for exposure. Several features in the HETGS
image are evident: (1) a central, bright point source, (2) a
circumnuclear diffuse reflection component, (3) a soft X-ray plume
coincident with the known ionization cone, and (4) several off-nuclear
point sources.\footnote{We note that there may be another diffuse
emission component arising from unresolved point sources and/or
ionized gas associated with the optical star-forming disk. However,
this X-ray emission is diluted by the bright reflection halo that
extends out to large radii ($>$ 1\arcmin\ from the nucleus).} The
point-source distribution is elongated in the same direction as the
projected major axis of the optical disk (a position angle of about
$-45^{\circ}$ East of North), with the majority of the sources
concentrated towards the nucleus. We see a prominent division of the
circumnuclear reflection halo in Figure~\ref{fig:asmooth} into hard
(blue) and soft (red) X-ray 

\noindent emission which appears to be perpendicular
to the optical disk of the galaxy. This division is indicative of
intrinsic absorption and implies that there is a large gas and dust
reservoir lying in the plane of the galaxy near the nucleus. This is
consistent with the large absorption suggested from infrared
observations of the nuclear region \citep{Ruiz2000}.

The first three features of Circinus are discussed in
\citet[][b]{Sambruna2001a}. In the following analysis, we focus on the 
zeroth-order imaging and moderate-resolution spectroscopy of the
fourth component. We used the {\tt CIAO} source detection 
tool {\tt wavdetect} \citep{Freeman2001} to identify all point
sources above a threshold significance of 10$^{-7}$.
A total of 16 sources were found. Table~\ref{sources} lists these
sources along with their 0.5--10~keV fluxes, luminosities and spectral
characteristics (see $\S$\ref{spectra}). Point-source counts and
spectra were extracted using the 95\% encircled-energy radii at 1.5~keV
(4 pixels on-axis and up to 7--8 pixels 2\arcmin\
off-axis)\footnote{E.~D.~Feigelson 2001, private communication; see also 
http://asc.harvard.edu/cal/Links/Acis/acis/Cal\_prods/psf/analysis.html}
for all sources except CXOU~J141309.2$-$652017, CXOU~J141310.0$-$652021,
CXOU~J141310.3$-$652017, and CXOU~J141310.4$-$652022. For these
sources, the 90\% encircled-energy radii were used to limit the amount
of contamination from the diffuse circumnuclear emission component
\citep[for details on this diffuse component, see][]{Sambruna2001a}.
The background-subtracted counts measured from the 90\% and 95\%
encircled-energy radii were multiplied by 1.11 and 1.05 to obtain the
aperture-corrected count values given
in column 2 of Table~\ref{sources}. Since some sources are embedded in
diffuse emission well in excess of the typical 0.07~cts~pixel$^{-1}$
background for this observation, we extracted background counts using
circular or annular regions of 100--1000 pixels adjacent to or
surrounding the source extraction region. Our absorbed,
background-subtracted point-source detection limit is 7 counts,
corresponding to an absorbed flux of $F_{\rm
0.5-10~keV}=7\times10^{-15}$~erg~cm$^{-2}$ s$^{-1}$ or an
absorption-corrected luminosity of $L_{\rm 0.5-10~keV}=2\times
10^{37}$~erg~s$^{-1}$ (assuming an average power-law spectrum with
photon index $\Gamma=1.76$ and $N_{\rm
H}=1.5\times10^{22}$~cm$^{-2}$).

The optical nature of these serendipitous X-ray sources was assessed
using {\it HST} WFPC2 (F502N, F547M, F606W, F656N and F814W) archival
images; see Table~2 for details. The images were processed
through the Space Telescope Science Institute pipeline.\footnote{See
http://www.stsci.edu/documents/dhb/web/DHB.html.} When multiple
images in a given filter were available, the images were combined to
reject cosmic rays using the {\tt IRAF} task {\tt crrej}. The
automatic star-finding algorithm {\tt daofind} in {\tt DAOPHOT}
\citep{Stetson1987} was used to generate a preliminary list of stars
detected in each chip for each filter. We then performed aperture
photometry to determine each star's instrumental magnitude using
2-pixel radius apertures in the WF chips and 5-pixel radius apertures
in the PC chip. Aperture corrections to a 0\farcs5 radius were then
determined from bright stars on each chip.

\end{multicols}
\begin{deluxetable}{lrrrrrrrrl}
\tabletypesize{\small}
\rotate
\tablewidth{0pt}
\tablecaption{X-ray Sources in the ACIS-S3 Image of Circinus\label{sources}} 
\tablehead{
\colhead{(1)} & 
\colhead{(2)} & 
\colhead{(3)} & 
\colhead{(4)} & 
\colhead{(5)} & 
\colhead{(6)} & 
\colhead{(7)} & 
\colhead{(8)} & 
\colhead{(9)} &
\colhead{(10)} \\
\colhead{CXOU~J Source} & 
\colhead{Counts} & 
\colhead{$P_{\rm con}$} & 
\colhead{$m_{\rm V}$} & 
\colhead{$F_{\rm X}$} & 
\colhead{$L_{\rm X}$} & 
\colhead{log($F_{\rm X}/F_{V}$)} & 
\colhead{$N_{\rm H}$} & 
\colhead{$\Gamma$} & 
\colhead{Comments} \\
}
\startdata
141257.8$-$651425 &   38.0 $\pm$ 10.9 & 97.3 & $>20.7$ (DSS)       & 0.040\tablenotemark{\dagger} & 0.116\tablenotemark{\dagger}  & $> 0.25$ &                \nodata &                \nodata & \\
141303.3$-$652043 &   17.4 $\pm$  6.5 & 42.6 & $ 22.1$ ({\it HST}) & 0.019\tablenotemark{\dagger} & 0.053\tablenotemark{\dagger}  & $  0.49$ &                \nodata &                \nodata & \\
141305.5$-$652031 &   21.3 $\pm$  7.0 & 38.6 & $>25.3$ ({\it HST}) & 0.023\tablenotemark{\dagger} & 0.065\tablenotemark{\dagger}  & $> 1.85$ &                \nodata &                \nodata & \\  
141307.5$-$652106 &    7.7 $\pm$  3.9 &  8.4 & $ 22.2$ ({\it HST}) & 0.007\tablenotemark{\dagger} & 0.020\tablenotemark{\dagger}  & $  0.10$ &                \nodata &                \nodata & Variable (flare) \\
141309.2$-$652017 &   99.1 $\pm$ 12.0 & 43.8 & $>23.4$ ({\it HST}) & 0.10                         & 0.24                          & $> 1.73$ & $0.99$ ($<4.87$)       & $1.44^{+1.36}_{-0.93}$ & \\
141309.7$-$652211 &   11.9 $\pm$  4.6 & 29.8 & $>20.7$ (DSS)       & 0.011\tablenotemark{\dagger} & 0.032\tablenotemark{\dagger}  & $>-0.31$ &                \nodata &                \nodata & \\
141310.0$-$652021 & 6571.2 $\pm$ 81.4 & 85.6 & $\sim 19.0$ ({\it HST}) & 12.3                     & 23.3                          & \nodata  &                \nodata &                \nodata & Nucleus \\
141310.0$-$652044 (CG~X-2) & 1374.5 $\pm$ 37.1 & 76.0 & $ 22.2$ ({\it HST}) & 1.33                & 3.43                          & $  2.61$ & $1.14^{+0.29}_{-0.26}$ & $1.77^{+0.21}_{-0.22}$ & Long-term variable, 6.9~keV emission line, \\
                           &                   &      &                     &                     &                               &          &                        &                        & embedded in HII region ({\it HST}) \\
141310.1$-$651832 &   15.9 $\pm$  5.7 & 54.5 & $>20.9$ (DSS)       & 0.017\tablenotemark{\dagger} & 0.048\tablenotemark{\dagger}  & $>-0.04$ &                \nodata &                \nodata & \\ 
141310.1$-$652029 &  166.0 $\pm$ 15.0 & 93.2 & $>25.3$ ({\it HST}) & 0.17                         & 0.73                          & $> 2.28$ & $3.37^{+3.49}_{-2.95}$ & $2.03^{+1.14}_{-0.92}$ & \\
141310.3$-$652017 &  342.5 $\pm$ 20.6 & 85.1 & $>25.3$ ({\it HST}) & 0.26                         & 1.40                          & $> 1.94$ & $1.26^{+0.68}_{-0.65}$ & $2.72^{+0.51}_{-0.52}$ & \\
141310.4$-$652022 &  792.7 $\pm$ 29.7 &  2.3 & $>25.3$ ({\it HST}) & 1.08                         & 2.09                          & $> 3.08$ & $0.91^{+0.65}_{-0.43}$ & $0.68^{+0.31}_{-0.20}$ & Variable (rise) \\
141312.2$-$652007 &  186.0 $\pm$ 14.9 &  0.9 & $ 20.1$ (DSS)       & 0.21                         & 0.52                          & $  0.73$ & $1.64$ ($<4.22$)       & $1.37^{+0.80}_{-0.84}$ & Variable (flare) \\
141312.3$-$652013 (CG~X-1) & 1095.2 $\pm$ 34.2 &  0.0 & $>25.3$ ({\it HST}) & 0.90                & 3.71                          & $> 3.44$ & $1.26^{+0.28}_{-0.22}$ & $2.40^{+0.22}_{-0.19}$ & Variable (periodic) \\
141312.6$-$652052 &  119.2 $\pm$ 12.2 &  0.4 & $>25.3$ ({\it HST}) & 0.12                         & 0.33                          & $> 1.65$ & $1.43$ ($<5.26$)       & $1.63^{+1.61}_{-0.64}$ & Variable (flare) \\ 
141312.9$-$651957 &   27.7 $\pm$  7.1 & 16.8 & $>18.1$ (DSS)       & 0.029\tablenotemark{\dagger} & 0.084\tablenotemark{\dagger}  & $>-0.93$ &                \nodata &                \nodata & \\
\tableline
1\arcmin\ radius aperture &     
                     11760 $\pm$ 152  &  4.2 & & 10.0                         & 17.2  &          &                        &                        & \\
\enddata
\tablecomments{Column 1: Source name given as CXOU~JHHMMSS.S+DDMMSS. The 
final row lists the global properties of Circinus determined within a
1\arcmin aperture radius of the nucleus using software provided by
A.~Vikhlinin
(http://asc.harvard.edu/cgi-gen/contributed\_software.cgi). Column 2:
Background-subtracted, aperture-corrected 0.5--10.0~keV counts
accumulated over 67~ks. Aperture photometry was performed using 90\%
and 95\% encircled-energy radii for 1.5~keV, and individual background
regions were selected adjacent to each source as noted in
$\S$\ref{detection}. The standard deviation for the source and
background counts are computed following the method of
\citet{Gehrels1986} and are then combined following the ``numerical
method'' described in $\S$1.7.3 of \citet{Lyons1991}. Column 3: The
probability that the source count rate is consistent with a constant
rate, as derived from the Kolomogrov-Smirnov test (in percent). Column
4: Visual magnitudes or lower limits as determined from {\it HST} or
DSS images (see $\S$\ref{detection}). Column 5: Observed 0.5--10.0~keV
fluxes in units of 10$^{-12}$~erg~cm$^{-2}$~s$^{-1}$ from the best-fit
models to the ACIS spectra. Fluxes for sources denoted by a $\dagger$
were calculated assuming an average power-law spectrum with $N_{\rm H}
= 1.5\times10^{22}$ cm$^{-2}$ and photon index $\Gamma=1.76$ as
determined from the best-fit values of sources with more than 90
counts. Column 6: Absorption-corrected 0.5--10.0~keV luminosities in
units of 10$^{39}$~erg~s$^{-1}$ from the best-fit models to the ACIS
spectra. Luminosities for sources denoted by a $\dagger$ were
calculated with the same assumptions as outlined in Column 5. Column
7: Ratio of the 0.5--10.0~keV X-ray flux to the $V$ magnitude optical
flux. Column 8: Neutral hydrogen absorption column density in units of
$10^{22}$ cm$^{-2}$ as determined from the best-fit models to the ACIS
spectra. Also listed are the 90\% confidence errors calculated for one
parameter of interest ($\Delta\chi^2 = 2.7$). Column 9: Power-law
photon index $\Gamma$ as determined from the best-fit models to the
ACIS spectra. Also listed are the 90\% confidence errors calculated
for one parameter of interest ($\Delta\chi^2 = 2.7$). Column 10:
Comments on X-ray variability and optical identifications.}
\end{deluxetable}
\begin{multicols}{2}

{\footnotesize
\begin{center}
{\sc Table 2} \\
\vspace{0.02in}
{\sc {\it HST} Archival Images of Circinus} \\
\vspace{0.06in}
\begin{tabular}{cccc}
\hline \hline \vspace{-0.06in} \\
{(1)} & {(2)} & {(3)} & {(4)} \\ 
{Filter\tablenotemark{\dagger}} & {Exp. Time (s)} & {$m_{\rm lim}$} & 
{Obs. Date} \\ [0.1cm]
\hline  \vspace{-0.08in} \\
WFPC2 F502N & 900, 900 & 23.3 & 1999 Apr. 10 \\ [0.05cm]
WFPC2 F547M &       60 & 23.0 & 1999 Apr. 10 \\ [0.05cm]
WFPC2 F606W & 200, 400 & 25.3 & 1996 Aug. 11 \\ [0.05cm]
WFPC2 F656N & 800, 800 & 22.7 & 1999 Apr. 10 \\ [0.05cm]
WFPC2 F814W &       40 & 22.8 & 1999 Apr. 10 \\ [0.05cm]
\hline \\
\end{tabular}
\vspace{0.1in}
\parbox{2.80in}{ 
\small\baselineskip 9pt
\footnotesize
\indent $^{\dagger}$  {The WFPC2 filter name denotes the
central filter wavelength in nanometers and the filter width
(N=narrow, M=medium, W=wide).}
}
\end{center}}
\setcounter{table}{2}
\vspace{-0.06in}

We aligned the {\it HST} images to the {\it Hipparcos}/Tycho
astrometric reference frame using three moderately bright stars from
the Tycho catalog \citep{Hog2000} to provide absolute astrometry to
$\approx$ 0\farcs4. The F606W image had approximately the
same aim point as the other WFPC2 images, but it was rotated by
130$^{\circ}$ such that most (but not all) of a 2\arcmin\ field
centered on Circinus was imaged by WFPC2. We filled the slight optical
coverage gap using 0\farcs7 resolution 5100 \AA\ and 7000 \AA\
continuum images taken with the Superb-Seeing Imager on the New
Technology Telescope (NTT) in 1994 \citep[Figures 1 and 2
of][]{Marconi1994}. The {\it Chandra} zeroth-order field-of-view is
significantly larger than either the {\it HST} or NTT images, so the
optical nature of the few sources farthest from the center of Circinus
was evaluated using the Digitized Sky Survey
\citep[DSS;][]{Lasker1990}. Unfortunately, the coarse resolution (FWHM
$\approx$ 3\arcsec) and shallow depth (limiting $m_{\rm B}\approx 20$)
of the DSS image in this region of the sky are not nearly as
constraining as the {\it HST} images. The NTT and DSS images were
aligned with the {\it HST} images using ten bright, isolated point
sources in the field. The astrometric accuracy of each image is
$\approx$ 1\farcs0.

The only obvious optical counterpart to any of the X-ray sources is
the nucleus of Circinus, which is offset from the X-ray nucleus
centroid by 0\farcs7. Given the slight uncertainty in the astrometry
of the {\it HST} images, this value is consistent with the typical
astrometric accuracy for reprocessed {\it Chandra} data (i.e.,
$\la$~$0\farcs6$).\footnote{See
http://asc.harvard.edu/mta/ASPECT/cel\_loc/cel\_loc.html.} Aligning
the images based on this shift, we found that three of the off-nuclear
X-ray sources are coincident with {\it HST} optical counterparts to
within 0\farcs5: an $m_{\rm V}=22.1$ counterpart with
CXOU~J141303.3$-$652043, an $m_{\rm V}=22.2$ counterpart with
CXOU~J141307.5$-$652106, and an $m_{\rm V}=22.8$ counterpart with
CXOU~J141310.0$-$652044. Johnson $V$ magnitudes for these sources were
calculated using {\tt SYNPHOT} assuming a spectral slope determined
from their $m_{\rm F814}-m_{\rm F547}$ color. For X-ray sources
without {\it HST} detections, $V$ magnitude limits were calculated
using {\tt SYNPHOT} assuming a spectral slope determined from the
average $m_{\rm F814}-m_{\rm F547}$ color for sources in the field
(i.e., $m_{\rm F814}-m_{\rm F547}\approx 1.4$). Several of the
detected X-ray sources, however, fall outside of the range of the {\it
HST} observations. Inspection of the NTT 5100~\AA\ continuum and DSS
images reveals an $m_{\rm V}=20.1$ counterpart at a distance of
0\farcs8 from another source (CXOU~J141312.2$-$652007). Johnson $V$
\centerline{\hglue0.5cm{\includegraphics[width=4cm]{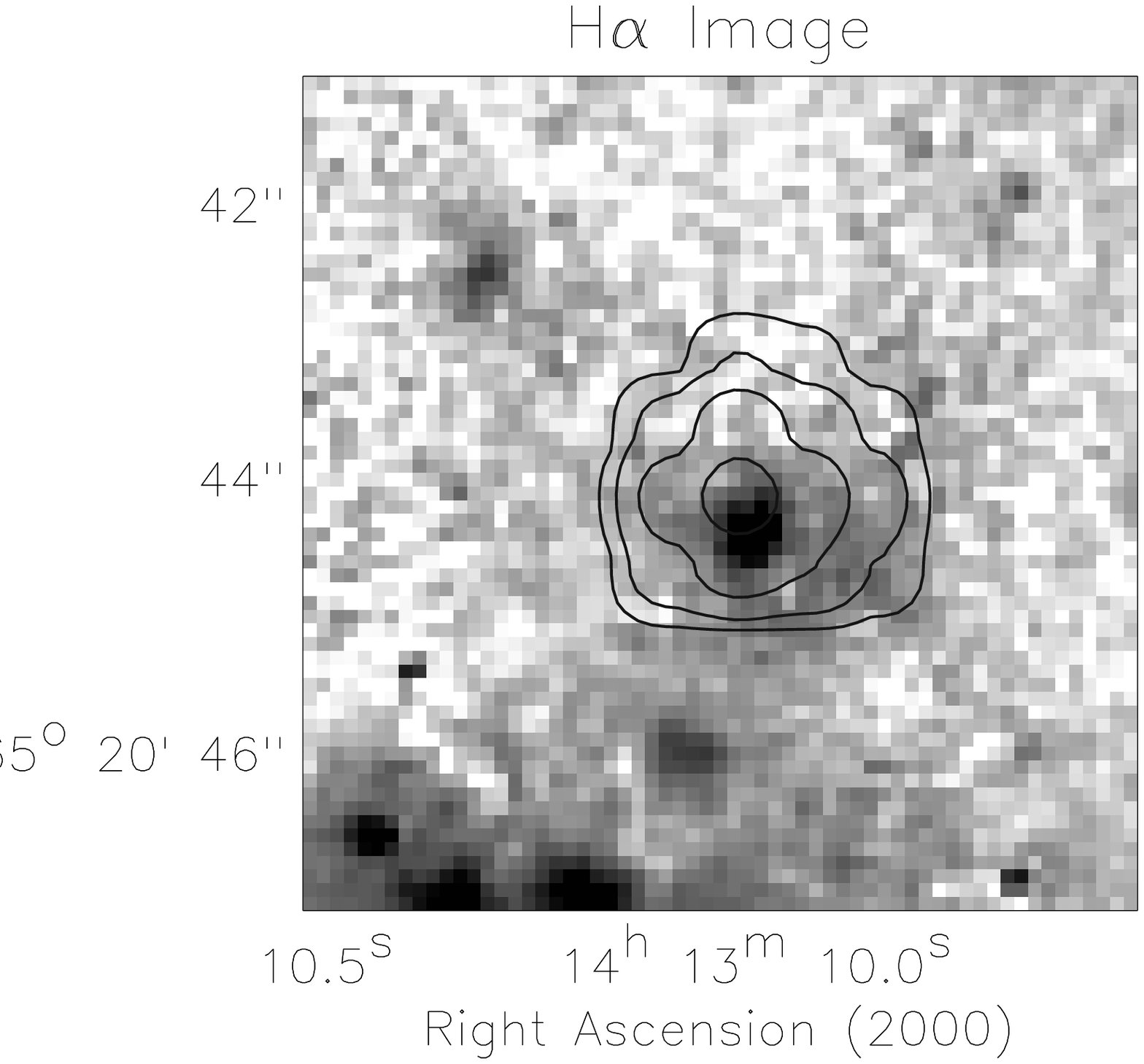}} \hfill
\hglue0.5cm{\includegraphics[width=4cm]{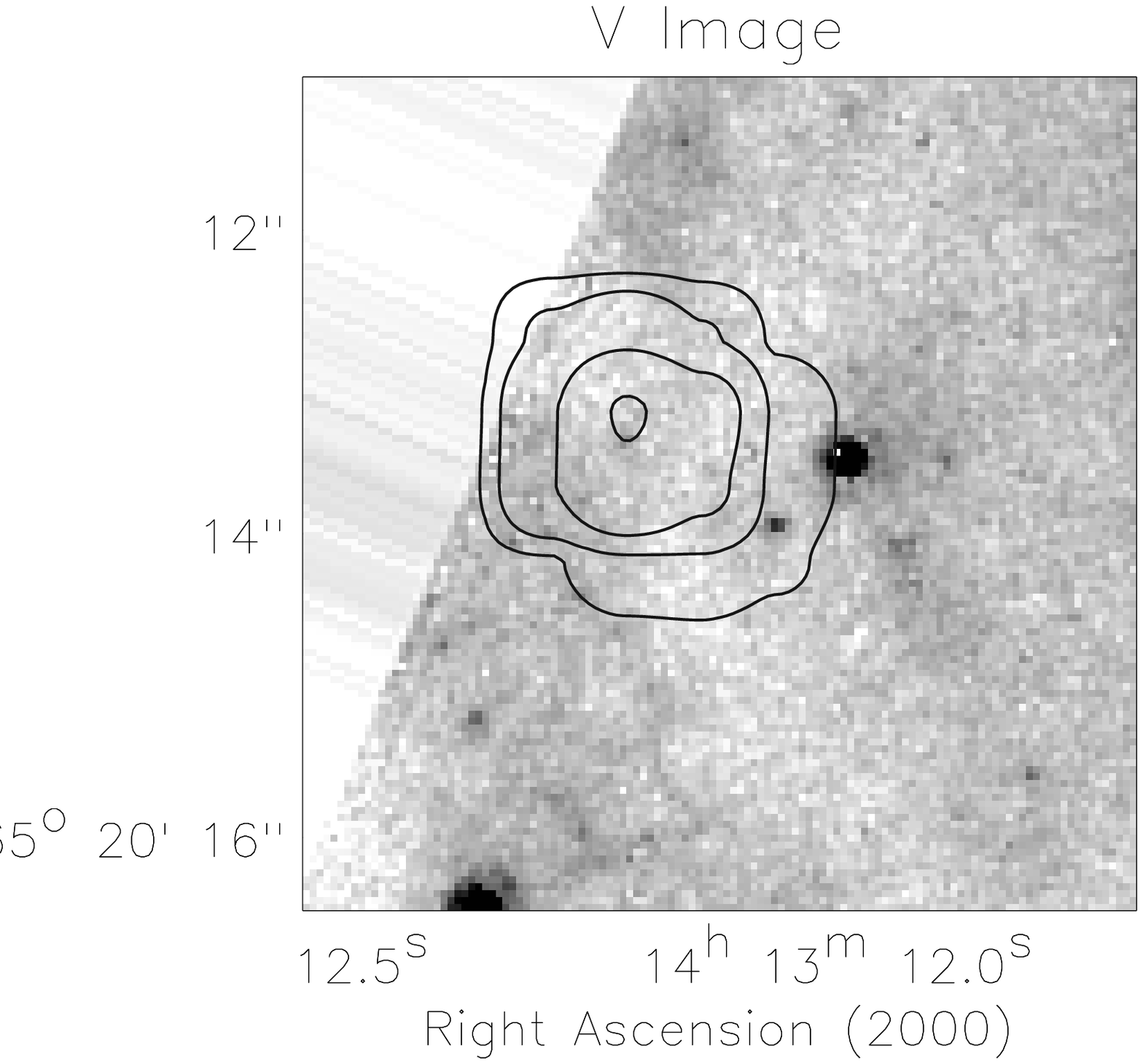}}}
\figcaption[fig2a.eps]{
HETGS zeroth-order contours of the two brightest off-nuclear sources,
CG~X-2 (left) and CG~X-1 (right),
overlaid on {\it HST} F656 WF4 and F606 PC chip images,
respectively. The lowest contour level is 4 counts, with each
successive contour increasing by a factor of 4. CG~X-2 
coincides with an $m_{\rm V}=22.2$ companion embedded in a probable 
\ion{H}{2} region, as indicated by the diffuse emission surrounding the 
optical point source in the {\it HST} $\mathrm{H}\alpha$ image. 
CG~X-1 has no optical counterpart to $m_{\rm V}>25.3$.
\label{fig:overlay}} 
\vspace{0.2in}

\noindent
magnitudes and lower limits for
sources covered by the NTT and DSS images were calculated using {\tt
SYNPHOT} assuming a spectral slope determined from the average $B-R$
color for sources in the field (i.e., $B-R \approx
1.25$).\footnote{Johnson $B$ and $R$ magnitudes were derived from the
United States Naval Observatory A2.0 Catalog $O$ and $E$ plate
magnitudes using using $B = O - 0.119(O-E)$ and $R = E$
\citep{Evans1989}.}

In Figure~\ref{fig:overlay}, we show contours of the two brightest
off-nuclear X-ray sources superimposed on {\it HST} postage-stamp
images, both of which are integral to later discussions. There is no
detectable counterpart at the X-ray position of CXOU
J141312.3$-$652013 (hereafter CG~X-1). CXOU~J141310.0$-$652044
(hereafter CG~X-2) coincides with an $m_{\rm V}=22.2$ companion
embedded in a probable \ion{H}{2} region, as indicated by the diffuse
emission surrounding the optical point source in the {\it HST}
$\mathrm{H}\alpha$ image in Figure~\ref{fig:overlay}. The bright,
compact $\mathrm{H}\alpha$ emission ($m_{\rm H\alpha}=17.9$) from this
source may originate in either a compact \ion{H}{2} region or a young
SNR. In general, we found that many of the serendipitous X-ray sources
appear to be coincident with optical star-forming regions:
CXOU~J141310.3$-$652017 with the inner 40~pc star-forming ring;
CXOU~J141310.1$-$652029 with the outer 200~pc star-forming ring;
CG~X-2 with an inter-arm $\mathrm{H}\alpha$ complex; and
CXOU~J141309.2$-$652017, CXOU~J141310.4$-$652022, CXOU
J141312.2$-$652007 and CG~X-1 with the spiral arms.

After assessing the optical properties of the X-ray point sources, we
used X-ray-to-optical flux ratios to evaluate different X-ray emission
mechanisms. For instance, \citet{Maccacaro1988} found that normal
stars have $-4.3<\log(F_{\rm X}/F_{\rm V})<-0.3$ while background
galaxies and AGN have $-1.5<\log(F_{\rm X}/F_{\rm
V})<1.9$.\footnote{The typical X-ray-to-optical flux ratios derived by 
\citet{Maccacaro1988} for individual source types were based on X-ray
observations in the 0.3--3.5~keV band with low neutral hydrogen column
densities. However, all of the Circinus off-nuclear sources exhibit
spectral cutoffs below $\sim$~1~keV (see $\S$\ref{spectra}).
Therefore, the flux ratios listed here have been corrected for the
Galactic absorption column and visual extinction given in
$\S$\ref{intro}. The flux ratios have also been converted from
0.3--3.5~keV to 0.5--10.0~keV assuming the average power-law spectrum
measured for sources in the Circinus field. Large spectral deviations
from our adopted model could result in $\log(F_{\rm X}/F_{\rm V})$
differences of up to $\sim$ 0.5--1.0.} RS CVn stars, XBs and SNRs are
harder to characterize because their $log(F_{\rm X}/F_{\rm V})$ ratios
can range from values close to those of normal stars (RS CVn stars) up
to ratios greater than 3.0 \citep[XBs; see][]{Maccacaro1988}. This
large range in X-ray-to-optical flux ratio leaves some ambiguity in
the classification scheme for sources with $\log(F_{\rm X}/F_{\rm
V})<1.9$, but the scheme is fairly clear cut for sources with larger
ratios. In the case of Circinus, the off-nuclear sources have
$\log(F_{\rm X}/F_{\rm V})$ ranging from $-$0.9 to 3.4 (see column 7
of Table~\ref{sources}). We find that five off-nuclear point sources
are unambigiously classified as XBs or SNRs. The remaining sources
could be similarly classified, but the X-ray-to-optical flux ratios
alone cannot exclude the possibility that eight sources could be
background galaxies or AGN and two sources could be normal stars. We
address the probability of finding foreground or background sources
coincident with Circinus below.

Nearly all of the detected X-ray point sources are coincident with the
disk of Circinus and lie within 2\arcmin\ of the nucleus. However,
Circinus lies behind a crowded stellar region at low Galactic latitude
($b=-3\fdg8$), so the chance superposition of Galactic X-ray sources
is a valid concern. Furthermore, Circinus has a Galactic longitude of
$l=311\fdg3$, so our line-of-sight to Circinus intersects a sizable
fraction of the Galaxy. To estimate the number of foreground or
background sources expected within the central region of Circinus, we
used a deep {\it Chandra} observation of the Galactic Ridge
($l=30^{\circ}$, $b=0^{\circ}$; K. Ebisawa et al., in preparation).
This observation is at a similar Galactic latitude and has a limiting
X-ray flux of $F_{\rm 2-10~keV}=3\times
10^{-15}$~erg~cm$^{-2}$~s$^{-1}$. Since Circinus and the Galactic
Ridge have comparable angular distances from the Galactic Center, they
should also have comparable source-count distributions. In the
2--10~keV band, the Circinus zeroth-order image has a flux threshold
of $F_{\rm 2-10~keV} = 6\times10^{-15}$~erg~cm$^{-2}$~s$^{-1}$
assuming $\Gamma=1.76$ and $N_{\rm H}=1.5\times10^{22}$~cm$^{-2}$.
From the Galactic Ridge measurements, we expect to detect
$\approx$~300 sources per deg$^{2}$ in a 67~ks {\it Chandra} HETGS
observation, corresponding to $\approx$ 1.0 foreground or background
sources within a 2\arcmin\ radius of Circinus at our detection
threshold (see Figure~\ref{fig:asmooth}). Therefore, most of the
sources within 2\arcmin\ of the nucleus should be associated with
Circinus.

Alternatively, we could ask how many sources we would expect to find
in the field outside the disk of Circinus (i.e., $>$ 2\arcmin\ from
the Circinus nucleus)? If we exclude Circinus from the ACIS S3
chip subarray, we expect to find $\approx$~2 sources in the remaining
area ($\approx$~27~arcmin$^{2}$) after correcting for vignetting
losses. Only one source is detected within this region
(CXOU~J141257.8$-$651425), indicating that our use of the Galactic
Ridge number counts is appropriate and perhaps somewhat pessimistic.

Finally, we note that sources significantly brighter than our limiting
flux have much more stringent limits. Since the {\it Chandra} Galactic
Ridge number counts are poorly sampled above $F_{\rm 2-10~keV} =
1\times10^{-13}$~erg~cm$^{-2}$~s$^{-1}$, we combined them with the
results of the {\it ASCA} Galactic Plane Survey \citep{Sugizaki1999}
to estimate that there is a $\la$ 0.06\% chance that either of the
bright off-nuclear sources shown in Figure~\ref{fig:overlay} (i.e.,
CG~X-1 and CG~X-2) are foreground or background X-ray sources.

\section{Timing Analysis}\label{timing}

\subsection{Short-Term Variability}\label{short-term}

The {\it Chandra} observation of Circinus was of sufficient duration
to evaluate short-term timing characteristics for many of the
off-nuclear sources. To determine objectively the existence of
significant variations in the count rate, we used the
Kolmogorov-Smirnov (KS) statistic to test the null hypothesis that
each source plus background rate was constant over the duration of the
exposure.\footnote{The data were not binned for the KS test. The
minimum time resolution was the frame time.} Thus a source with a low
KS probability ($P_{\rm con}$) has a high probability of being
variable ($1-P_{\rm con}$). Since the timing gap in the observation
could lead to inaccurate variability estimates, we modified the KS
test to exclude the 1.3~ks gap between the continuous 6.9~ks and
60.2~ks segments of the observation. We find that sources 
\centerline{{\hglue-0.5cm{\includegraphics[width=9.0cm]{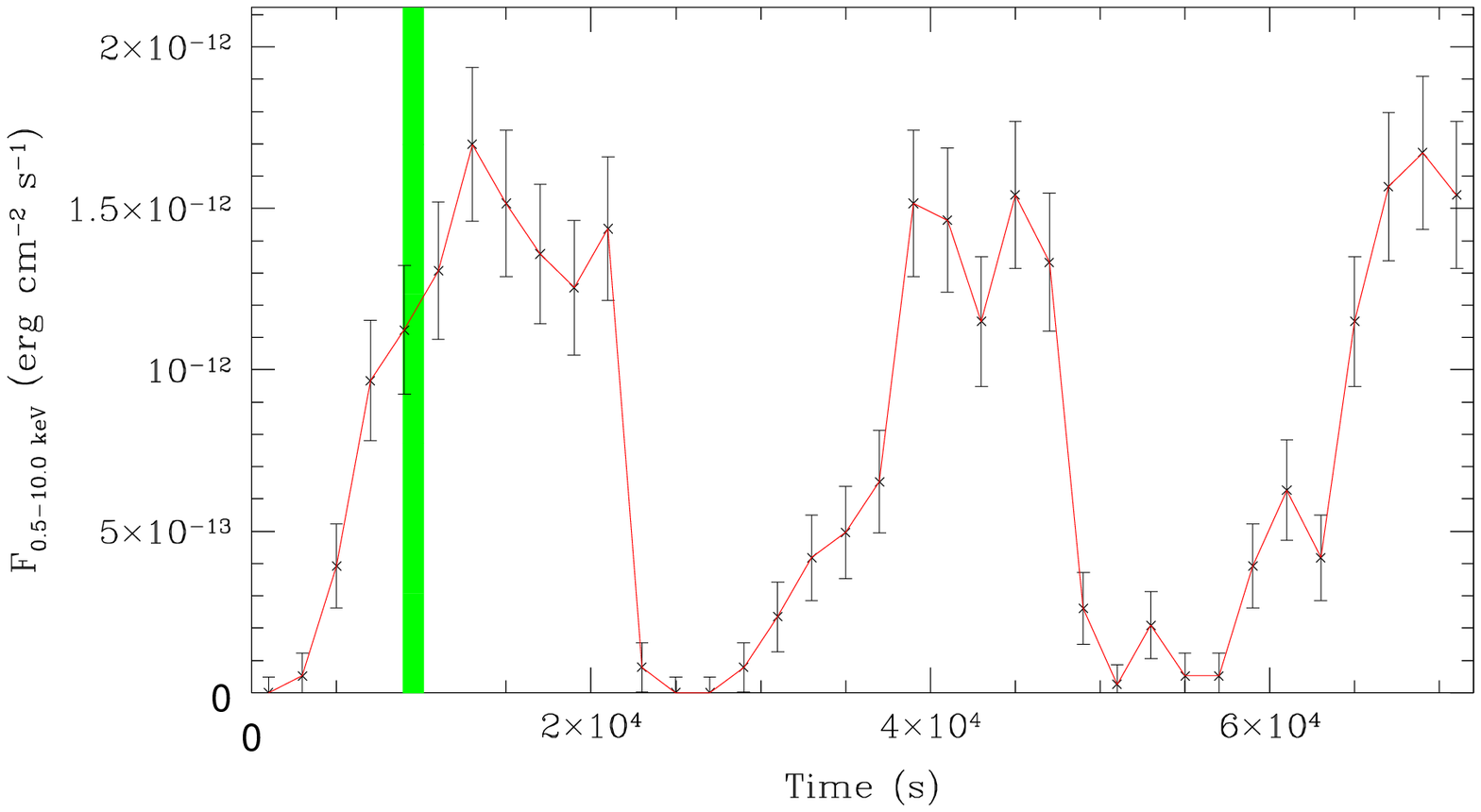}}}}
\vspace{-1.65in}
\figcaption[fig3.eps]{
Light curve of CG~X-1 over the entire duration of the observation
(i.e., 6.9~ks + 1.3~ks gap + 60.2~ks) binned in 2~ks intervals. The
period is $27.0\pm0.7$~ks. In that time, the flux varies by more than
a factor of 20. The 1.3~ks gap (shaded band) was interpolated using
the count rates found from averaging the 1~ks data segments before and
after the gap. The errors bars have been calculated following the
method of \citet{Gehrels1986}.
\label{fig:periodic}} 
\vspace{0.2in}
\noindent CXOU~J141312.2$-$652007, CG~X-1, and CXOU J141312.6$-$652052 
varied at the $>$ 99\% confidence level, while sources
CXOU~J141307.5$-$652106 and CXOU~J141310.4$-$652022 were less
convincing, varying at the $\approx$ 92\% and $\approx$ 98\%
confidence levels, respectively. Although CXOU~J141307.5$-$652106 only
has a KS probability of 92\% and 7.7 background-subtracted counts, all
of these counts were detected in a contiguous 20~ks segment of the
observation, indicating it is a strong flaring source. In total, three
of these short-term variable sources appeared to flare on $\sim$ 20~ks
timescales, while one exhibited a steady increase in flux over the
course of the observation and another displayed periodic behavior (see
below). The short-term KS statistic results for all 16 {\it Chandra}
sources and the nature of variability observed are summarized in
columns 3 and 10 of Table~\ref{sources}, respectively.

Many of our serendipitous X-ray sources are likely to be XBs (see
$\S$\ref{detection}). Since a significant fraction of XBs are
X-ray pulsars with pulse periods ranging from 0.01--1,000~s
\citep[e.g.,][]{White1995}, we generated power spectra for all of the
Circinus X-ray sources to search for pulsations. These power spectra
also allowed us to search for other periodic phenomena (e.g., eclipses
or regular X-ray bursts). Only one source, CG~X-1, was found to
display periodic behavior above the frame-time limit of 2.1~s.
Figure~\ref{fig:periodic} presents the light curve of CG~X-1, which
shows periodic flux variability by at least a factor of 20. Because
the 1.3~ks gap could introduce a small artificial dip in the light
curve, we interpolated the flux values during this time using the
count rates found from the 1~ks data segments before and after the
gap. We used the period-fitting routine of
\citet{Stetson1996} to quantify the period and its one-sigma error as
$7.5\pm0.2$~hr. To study the light curve in more detail, we folded it
about its period as shown in Figure~\ref{fig:folded} to increase the
signal-to-noise. The flux of the source exhibits a gradual rise early
in its phase ($\sim20$\% of cycle), then flattens out near the peak
flux ($\sim40$\% of cycle), and abruptly fades to a minimum near
zero ($\sim40$\% of cycle) to close out the period. As the flux
gradually increases in Figure~\ref{fig:folded} the spectrum hardens
briefly, suggesting that the emission during this portion of the
period may be partially absorbed at soft energies.

The simplest explanation for the variability of CG~X-1 is an eclipse
by an evolving binary companion, although a pure 
\centerline{{\hglue-0.5cm{\includegraphics[width=9.0cm]{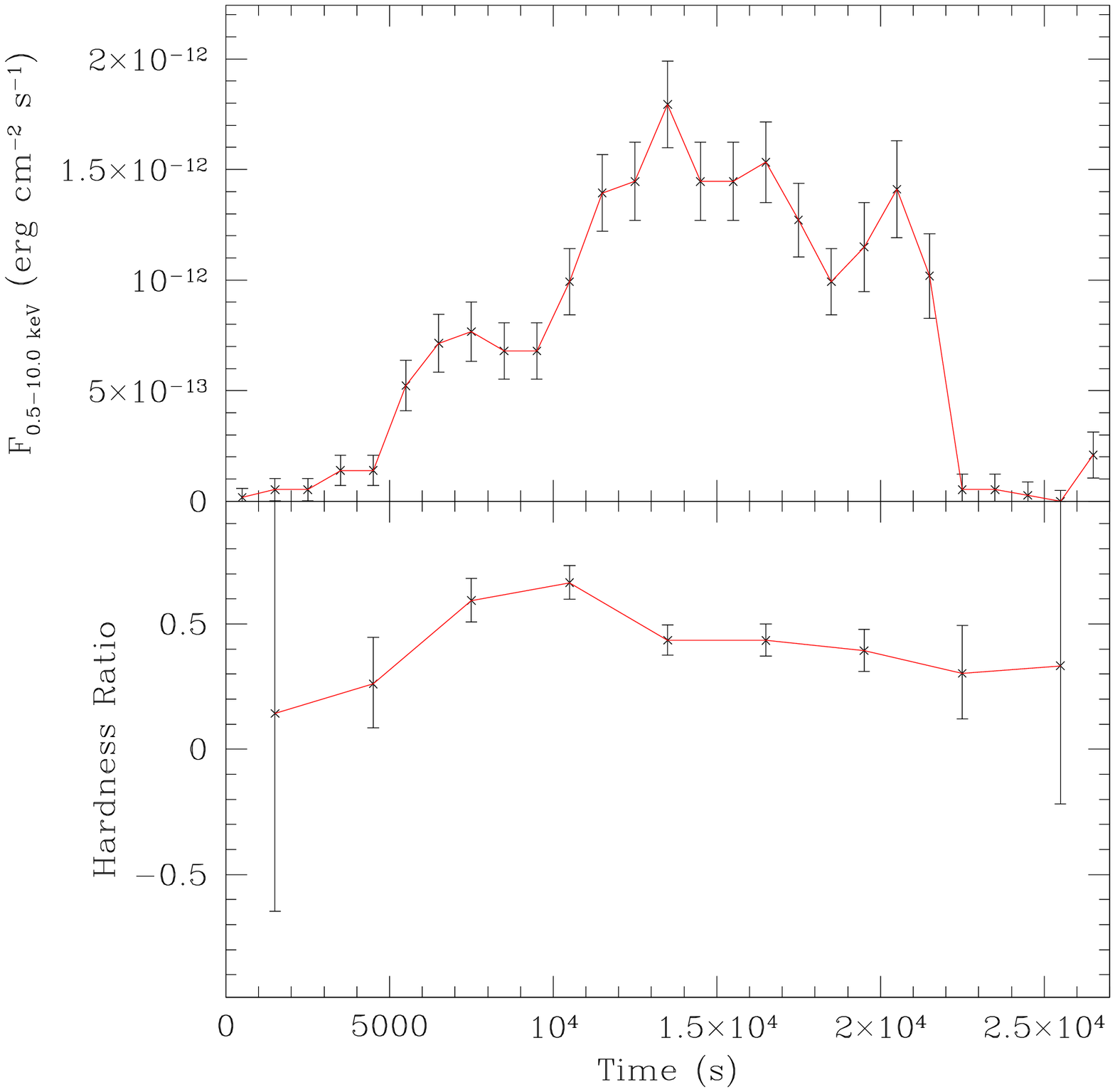}}}}
\figcaption[fig4.eps]{
(Top) Light curve of CG~X-1 folded about a period of 27.0~ks and
binned in 1~ks intervals. The error bars have been calculated
following the method of \citet{Gehrels1986} assuming only photon
noise. (Bottom) Hardness ratio (HR) of CG~X-1 folded about a period of
27.0~ks and binned in 3~ks intervals. HR = (H $-$ S)/(H $+$ S) where H
= 2--10~keV counts and S = 0.5--2~keV counts. The errors bars for this
quantity have been calculated following the ``numerical method''
described in $\S$1.7.3 of \citet{Lyons1991}. Prior to folding the flux
and HR light curves, the 1.3~ks gap was interpolated using the count
rates found from averaging the 1~ks data segments before and after the
gap.
\label{fig:folded}}
\vspace{0.2in}
\noindent eclipse model has
difficulty explaining certain aspects of the light curve. For
instance, the fraction of the period spent in eclipse is related to
the size of the Roche Lobe of the binary companion and hence to the
companion-to-compact-object mass ratio. We find that the eclipse
fraction of CG~X-1 is $\approx$ 40\%, implying a mass ratio of
$\approx$ 30 \citep[see the Appendix of][]{Pringle1985}. Such a large
mass ratio is difficult to reconcile with other observed quantities.
For instance, if CG~X-1 resides in the disk of Circinus at a distance
of 3.8 Mpc (see $\S$\ref{detection}), then its average and peak X-ray
luminosities are $3.7\times10^{39}$~erg~s$^{-1}$ and
$6.6\times10^{39}$~erg~s$^{-1}$. The peak luminosity implies a
compact-object mass of $M \ga 50 M_{\odot}$ if it is radiating at or
below its Eddington luminosity. The mass limit for the compact object
and the mass ratio predicted by a simple eclipse model are
incompatible with the standard model of stellar evolution (i.e., a
companion mass of $\ga 1500 M_{\odot}$ would be required). Therefore,
in the context of an eclipse model, the large eclipse fraction implies
that this binary system must either be experiencing strong dynamical
Roche Lobe overflow, or, alternatively, the X-ray emission region must
be at least partially occulted by some other means such as an
accretion disk around the compact object.

Another possible scenario for generating the large-amplitude periodic
variations seen from CG~X-1 is by modulation of the accretion rate
(e.g., from accretion-disk instabilities). One such example is the
extremely variable and luminous Galactic X-ray source, GRS~1915$+$105.
In its active state, the observed X-ray intensity of GRS~1915$+$105 is
found to vary dramatically on timescales ranging from seconds to days,
occasionally displaying quasi-periodic episodes
\citep[e.g.,][]{Greiner1996}. Some of these quasi-periodic episodes
coarsely resemble the light curve of CG~X-1 (although the timescales
are generally shorter for GRS~1915$+$105). Thus CG~X-1 could also
plausibly be an analog of GRS~1915$+$105 within Circinus.

We note that the asymmetric rise and decline in intensity of the
CG~X-1 light curve resembles those of AM~Her type
systems \citep[e.g.,][]{Warner1995}. This class of CV typically has
periods of 1--4~hr, intrinsic X-ray luminosities between
$10^{31}<L_{\rm 0.5-10~keV}<10^{33}$~erg~s$^{-1}$, and binary
companions occupying a narrow range of spectral types
\citep[M2V--M6V;][]{Cordova1995, Verbunt1995, Warner1995}. These
characteristics restrict the X-ray-to-optical flux ratio to $3<F_{\rm
X}/F_{\rm V}<300$. Given the X-ray flux associated with CG~X-1
(Table~\ref{sources}), we would expect an M star companion to have
$m_{\rm V} \approx$ 16--20, which is easily ruled out by the {\it HST}
limit. The {\it HST} limit requires that any M2V--M6V companion star
of CG~X-1 be at least 1.2 kpc away. The implied X-ray luminosity
($>10^{34}$~erg~s$^{-1}$) at this distance is at least 10 times larger
than the most luminous known AM~Her system. Furthermore, the period
(7.5~hr) is nearly two times longer than the longest period AM~Her
system known, and the X-ray spectrum is inconsistent with the
two-component model typical of these systems (see $\S$\ref{spectra}).
Based on these facts alone, a Galactic identification is doubtful.
Moreover, the observed X-ray flux of this object limits the
probability of it being a foreground source to $\la$~0.06\% (see
$\S$\ref{detection}). We also note that AM~Her systems in particular
comprise only a small fraction of all Galactic X-ray emitters
\citep[$\sim$~2\%; e.g.,][]{Motch1998}. Therefore the most promising
explanantion for CG~X-1 is an intermediate-mass black hole within
Circinus.

\subsection{Long-Term Variability}\label{long-term}

Circinus was also observed with the {\it ROSAT} HRI
($\approx$~5\arcsec\ FWHM; 0.1--2.4~keV) in 1995 September
\citep{Guainazzi1999}. With a total exposure time of only 4.2~ks, this
HRI observation is limited to source detections above $1.7 \times
10^{-14}$~erg~cm$^{-2}$~s$^{-1}$ (0.5--2.0~keV). CG~X-1 was detected
with a 0.5--2.0~keV flux of $(8.7 \pm 2.9)\times
10^{-14}$~erg~cm$^{-2}$~s$^{-1}$, a factor of $\sim$~2 lower than the
average 0.5--2.0~keV flux measured with {\it Chandra} ($1.7\times
10^{-13}$~erg~cm$^{-2}$~s$^{-1}$), but still well within the range of
intensity variations observed by {\it Chandra}. CG~X-2, on the other
hand, is completely absent from the HRI image, corresponding to a 95\%
confidence upper limit of $3.3\times 10^{-14}$~erg~cm$^{-2}$~s$^{-1}$
\citep[using the method of][]{Kraft1991} in the 0.5--2.0~keV band.
This is a factor of at least 11 smaller than the 0.5--2.0~keV flux of
$(4.1 \pm 0.1)\times10^{-13}$~erg~cm$^{-2}$~s$^{-1}$ measured with
{\it Chandra}. CG~X-2 is coincident with an \ion{H}{2} region, and the
{\it Chandra} radial profile is consistent with that of a point
source, indicating that this source could be either a moderately young
SNR or transient XB in Circinus. Observations taken with the
Australian Telescope Compact Array (ATCA) in 1995 with a beam size of
$\theta_{\rm Beam}=0\farcs9\times0\farcs8$ show an unresolved radio
source ($S_{\rm 3~cm}\sim 1.2$~mJy, $\alpha_{\rm J2000}=14^{\rm
h}$13$^{\rm m}$10\fs0 $\delta_{\rm
J2000}=-65^{\circ}$20\arcmin44\arcsec; M.~Elmouttie, private
communication) coincident to within $\la$~0\farcs6 of the X-ray
source, while none of the other radio point sources coincide with
X-ray sources \citep[see Figure 8 of][]{Elmouttie1998}. Given that
there are only a few significant radio point sources in the Circinus
field at a wavelength of 3~cm, the probability of this pair being a
random alignment is $<$ 0.1\%. The combination of bright, point-like
radio, optical, and X-ray emission strongly suggests that this object
is a young SNR or an \ion{H}{2} region. We can rule out radio emission
from a bright \ion{H}{2} region, however, since the observed flux at
3~cm is approximately two orders of magnitude higher than
expected.\footnote{We assume that a typical bright \ion{H}{2} region
has an electron temperature of $10^{4}$~K, electron and proton
densities of $10^{2}$~cm$^{-3}$, a size of less than 10~pc (for
details see Osterbrock 1989).} A young SNR classification for CG~X-2
is further supported by our spectral fitting in $\S$\ref{zero}. The
detection of a moderately strong radio source in 1995 and the lack of
a bright X-ray source in 1996 is curious but not implausible given
that the X-ray and radio emission are likely to result from physically
distinct and separate emission regions within a young SNR
\citep[e.g.,][]{Schlegel1995}. Ongoing optical monitoring programs of
Circinus over the past eight years have failed to detect any bright
supernovae down to $m_{\rm V}\approx 15$ (R. Evans 2000, private
communication; A. Williams 2000, private communication). Thus, if
CG~X-2 exploded after 1993, it must have been optically faint.

\section{Spectra}\label{spectra}

\subsection{Zeroth-order}\label{zero}

The ACIS zeroth-order spectra of the off-nuclear sources were
extracted for sources with more than $\approx$~90 counts using either
the 90\% or 95\% encircled-energy extraction radii as described in
$\S$\ref{detection}. Circinus was placed near the node 0/1 boundary on
the S3 chip, so several sources including the nucleus were dithered
between nodes 0 and 1. Since the variation in the energy response of
the S3 chip across these nodes is large, we chose to extract the
spectra and response matrices for each node separately and fit both
spectra with one model in {\tt XSPEC}. To assess the spectral nature of
these sources, we initially fitted the spectra with absorbed power-law
models.\footnote{In most cases, a Raymond-Smith thermal plasma model
\citep{Raymond1977} was equally acceptable.} We found that only one
off-nuclear source required more extensive spectral modeling (CG~X-2).
The best-fit models with 90\% confidence errors ($\Delta\chi^2 = 2.7$)
are listed in columns 8 and 9 of Table~\ref{sources}.

The range of spectral properties for these objects is quite diverse
(e.g., $\Gamma \sim$ 0.7--2.7). All of these off-nuclear sources
exhibit spectral cutoffs below $\sim$~1~keV that are best fitted with
column densities larger than the Galactic value (see $\S$\ref{intro}).
These large absorption values are consistent with the fact that many
of the off-nuclear X-ray sources appear to reside in star-forming
regions and spiral arms in Circinus (see $\S$\ref{detection}). The
spectra of the two brightest off-nuclear sources have adequate counts
to constrain physical models of emission and are discussed in more
detail below.

In Figure~\ref{fig:Src2-spectra}, we present the spectrum of the
periodic source CG~X-1. CG~X-1 happened to fall near the boundary of
node 0 and node 1, such that $\approx$~70\% of the photons from this
source were dithered onto node 0 and $\approx$~30\% onto node 1. We
show the node 0 spectrum in Figure~\ref{fig:Src2-spectra}. This
spectrum was initially fitted with a moderately soft power law with
$\Gamma=2.40^{+0.22}_{-0.19}$ and $N_{\rm
H}=(1.26^{+0.28}_{-0.22})\times10^{22}$~cm$^{-2}$ ($\chi^2=75.4$ for
59 degrees of freedom). Although this model proved acceptable,
\citet{Makishima2000} have shown that the X-ray spectra of
super-Eddington sources in spiral galaxies are often successfully
described by multicolor disk blackbody emission
\citep[MCD;][]{Mitsuda1984} arising from optically thick standard
accretion disks around black holes. We found that a MCD model with an
inner-disk temperature of $T_{\rm in}=1.35^{+0.13}_{-0.11}$~keV and
$N_{\rm H}=(6.39^{+1.57}_{-1.33})\times10^{21}$~cm$^{-2}$ worked
somewhat better than the power-law model ($\chi^2=61.0$ for 59 degrees
of freedom) and that the derived spectral parameters agree well with
typical innermost disk temperatures derived for other super-Eddington
sources \citep[i.e., $T_{\rm in}=1.1-1.8$~keV;][]{Makishima2000}.
Thus, we find that the spectral properties of this source are 
\centerline{\hglue-0.5cm{\includegraphics[height=8.5cm,angle=-90]{fig5a.eps}}}
\centerline{\hglue-0.5cm{\includegraphics[height=8.5cm,angle=-90]{fig5b.eps}}}
\figcaption[fig5a.eps]{
(Top) The upper panel shows the ACIS-S node 0 spectrum of CG~X-1, modeled
with a MCD model with an inner-disk radius temperature of $T_{\rm
in}=1.35^{+0.13}_{-0.11}$~keV and $N_{\rm
H}=(6.39^{+1.57}_{-1.33})\times10^{21}$~cm$^{-2}$. The lower panel shows
the residuals of the fit measured using the $\chi^2$ statistic. (Bottom)
Confidence contours between $T_{\rm in}$ and $N_{\rm H}$ for CG~X-1.
The contours represent the 68\%, 90\% and 99\% confidence levels for
two parameters of interest.
\label{fig:Src2-spectra}}
\vspace{0.2in}
\noindent fully
consistent with X-ray emission from an intermediate-mass black hole
candidate in the disk of Circinus. We also attempted to fit the
spectrum of CG~X-1 assuming the typical spectral model for an AM~Her
system (i.e., the most likely Galactic candidate) to remove any
remaining concerns about a possible Galactic nature of CG~X-1. Spectra
of AM~Her systems have substantial 6.4~keV iron $\mathrm{K}\alpha$
emission lines and continua comprised of a blackbody component with
$kT_{\rm BB}\sim20$ eV and a bremsstrahlung component $kT_{\rm
BR}\sim30$~keV
\citep[see $\S$6.4 of][]{Warner1995}. Such a model does not fit the
observed spectrum well and is rejected with $\gg$ 99\% confidence.

The other bright X-ray source, CG~X-2, was initially fitted with a
power law with $\Gamma=1.77^{+0.21}_{-0.22}$ and an absorption column
of $N_{\rm H}=(1.13^{+0.29}_{-0.26})\times10^{22}$~cm$^{-2}$ (see
Figure~\ref{fig:Src6-spectra1}). However, this fit was statistically
unacceptable with $\chi^2=106.5$ for 74 degrees of freedom and a
null-hypothesis probability of $P(\chi^2|\nu)=8.2\times10^{-3}$. The
spectrum shows large positive residuals around 6.9~keV. The addition
of a Gaussian emission line at $6.89^{+0.04}_{-0.09}$~keV with an
intrinsic line width of $190^{+57}_{-55}$~eV improved the fit
significantly ($\chi^2=63.2$ for 71 
\centerline{\hglue-0.5cm{\includegraphics[height=8.5cm,angle=-90]{fig6a.eps}}}
\centerline{\hglue-0.5cm{\includegraphics[height=8.5cm,angle=-90]{fig6b.eps}}}
\figcaption[fig6a.eps]{(Top) The upper panel shows the ACIS-S node 0 spectrum
of CG~X-2, modeled with a neutral hydrogen column of 
$N_{\rm H} = (1.13^{+0.29}_{-0.26})\times10^{22}$ cm$^{-2}$ and a
power law with $\Gamma = 1.77^{+0.21}_{-0.22}$. The lower panel shows
the residuals of the fit measured using the $\chi^2$ statistic. Note
the large $\chi^2$ residuals around 6.67--6.97~keV. (Bottom) The
confidence contours for fit parameters of a single Gaussian emission
line at 6.89~keV. The contours represent the 68\%, 90\% and 99\%
confidence levels for two parameters of interest.
\label{fig:Src6-spectra1}}
\vspace{0.2in}
\noindent degrees of freedom). The measured
equivalent width of the line is $1.59^{+0.41}_{-0.46}$~keV. We note
that this source is offset from the nucleus by 23\arcsec\ and totally
dominates over the faint background. Furthermore, the energy of this
line (6.89~keV) is quite different from the emission-line feature
found in the rest of the reflection halo \citep[$\approx$~6.4~keV, for
details see][]{Sambruna2001a} and is not associated with that
component.

To confirm that the central energy and line width of the 6.89~keV line
were not due to possible calibration problems, we measured the line
energies and widths of the 5.89~keV and 6.49~keV calibration lines in
a calibration dataset closest to our observation date (obs62017, 2000
June 18, $-120^{\circ}$~C). From the calibration data, we extracted a
large region on node 0 of the S3 chip near the position of CG~X-2 and
fit two Gaussian emission lines to the calibration lines using the
same response matrix as used for CG~X-2. The best-fit values were line
energies of $5.94^{+0.01}_{-0.01}$~keV and $6.54^{+0.01}_{-0.01}$~keV
and line widths of $53^{+03}_{-03}$~eV and $56^{+03}_{-03}$~eV,
respectively. Thus there appears to be a shift of $\approx$~50~eV in
the central line energy and uncertainty in the line width of
$\approx$~60~eV. The true central energy of the 6.89~keV line should
therefore be 6.84~keV. This 6.84~keV line is also broader than the
instrumental resolution of ACIS, suggesting that it is either
intrinsically broad or a blend of emission from two or more lines.

Given the central energy of the 6.84~keV emission line, it is likely
to be a blend of iron K-shell transitions at 6.67~keV (He-like line
complex) and 6.97~keV (H-like line). To test this hypothesis, we
modeled the spectrum of CG~X-2 using two Gaussian emission components
and an absorbed power law. We fixed the ratios of the He-like complex
and H-like line energies and line widths to be 0.957 and 1.0,
respectively, but we allowed the energy and width of the combination
to vary. The normalizations of the two lines were allowed to vary
independently. This model improved the $\chi^2$ to 57.5 for 70 degrees
of freedom and is better than the single Gaussian model at
$\approx$~99\% confidence using the $F$-test. The best-fit energies of
the iron lines are $6.74^{+0.03}_{-0.03}$~keV and
$7.05^{+0.03}_{-0.03}$~keV and are consistent with the line rest
energies after removing the 50~eV blueshift due to calibration errors
(see above). The intrinsic line width of each line is
$70^{+38}_{-27}$~eV and is consistent with unbroadened emission lines
(compare to line widths of calibration lines above). Allowing the
energies and widths of the lines to vary freely does not significantly
change the best-fit results, although the errors associated with each
parameter are much larger.

There are two scenarios that could explain such large equivalent width
iron K-shell line emission. We first explore the possibility that this
source is an X-ray binary where the strong emission lines must result
from fluorescence and resonant scattering either in a wind or a corona
around the accretion disk \citep[e.g., see $\S$1.6.2.3
of][]{White1995}. In this scenario, the direct line-of-sight to the
X-ray source must be blocked; otherwise the continuum emission from
the source would dilute the line emission and reduce its equivalent
width below what is observed. Scattering of the continuum by electrons
is relatively inefficient ($\approx$~1--5\%), so the intrinsic X-ray
luminosity of this source would have to be a factor of
$\approx$~20--100 times larger than observed. We also detect no
short-term variability for this source, so the direct continuum would
have to be blocked for more than 68~ks. Since the luminosity of this
source is already $L_{\rm 0.5-10.0~keV}=3.4\times10^{39}$~erg~s$^{-1}$
and exhibits no short-term variability, we consider an X-ray binary
scenario for CG~X-2 to be unlikely.

The second scenario to explain the X-ray line emission is that of a
young SNR (see $\S$\ref{long-term}). Within this scenario, the line
emission could be generated from either the initial forward shock that
passes through the outer layers of the star and propagates into the
circumstellar material or from the high-emissivity reverse shock that
develops at later times. The large X-ray luminosity of CG~X-2 is
comparable to those of other young ($\la$ 20 yr old) SNRs detected at
X-ray wavelengths \citep[e.g.,][]{Schlegel1995, Immler1998a, Fox2000,
Pooley2001}, adding merit to such a classification. To further test
this hypothesis, we fitted the spectrum with Raymond-Smith thermal
plasma and non-equilibrium ionization collisional plasma models
\citep{Borkowski1994}. Figure~\ref{fig:Src6-spectra2} shows the
best-fit model for an absorbed Raymond-Smith thermal plasma
[$kT=10.82^{+3.97}_{-2.52}$~keV, $Z=(1.35^{+0.84}_{-0.54}) Z_{\odot}$,
$N_{\rm H}=(9.21^{+1.70}_{-1.48})\times10^{21}$ cm$^{-2}$]. While this
model provides an acceptable fit to the data ($\chi^2=77.1$ for 73
degrees of freedom), it clearly underpredicts the width of the highly
ionized iron emission, leaving a gap between the H-like and He-like
iron lines around 6.8~keV. The non-equilibrium ionization model proved
no better at fitting the broad iron-line blend.

Only $\approx$ 12 young SNRs have X-ray detections at all, and when
adequate {\it ROSAT}, {\it ASCA} or {\it Chandra} spectra are
\centerline{\hglue-0.5cm{\includegraphics[height=8.5cm,angle=-90]{fig7.eps}}}
\figcaption[fig7.eps]{The upper panel shows the ACIS-S
node 0 spectrum of CG~X-2, modeled with a neutral hydrogen column of 
$N_{\rm H} = (9.21^{+1.70}_{-1.48})\times10^{22}$ cm$^{-2}$, a
Raymond-Smith thermal plasma model with 
$kT = 10.82^{+3.97}_{-2.52}$~keV and $Z=(1.35^{+0.84}_{-0.54}) Z_{\odot}$.
The lower panel shows the residuals of the fit measured using the
$\chi^2$ statistic. The emission-line blend centered at 6.89~keV is
poorly fit. \label{fig:Src6-spectra2}}
\vspace{0.2in}
\noindent available, most display high thermal temperatures ($kT=$ 5--10~keV)
similar to CG~X-2. Furthermore, only five young SNRs have been
observed with adequate sensitivity above 5~keV to detect any K-shell
iron emission. Three of these sources (SN~1986J in NGC~891, Houck et
al.~1998; SN~1993J in M81, Kohmura et al. 1994; SN~1998S in NGC~3877,
Pooley et al. 2001) appear to have substantial 6.8~keV iron-line
emission with equivalent widths of $\approx$ 0.3--2~keV. Thus the
spectrum of CG~X-2 appears to be consistent with the X-ray emission
seen from other young SNRs. The failure of simple physical models
(e.g., a Raymond-Smith model) to fully describe the iron K-shell
emission from CG~X-2 is intriguing, but not implausible considering
the scarcity of young, nearby SNRs and the lack of detailed X-ray
spectra. In addition to highly ionized iron K-shell emission, we might
expect to detect faint line emission from other highly ionized
elements such as sulphur and silicon. The spectrum of CG~X-2, however,
is too noisy to place reliable limits on such emission.

\subsection{HEG/MEG}\label{grating}

For the two brightest off-nuclear sources, CG~X-1 and CG~X-2, we
extracted MEG and HEG grating spectra (see Sambruna et al. 2001b for
details of the extraction procedure). We found that the
high-resolution spectra were consistent with the ACIS-S spectra, but
that the statistics were generally poor (less than 3$\sigma$ line
detections in the case of CG~X-2).

\section{Luminosity Function of Resolved Sources}\label{lf}

Luminosity functions (LFs) are often a useful tool for shedding
physical insight into the properties of source populations.
Figure~\ref{fig:lf} shows the cumulative distribution of the Circinus
off-nuclear sources as a function of luminosity. For comparison, we
also show three other galaxies spanning a large range of star
formation rates for which similar 0.5--10.0~keV source distributions
could be obtained: the Small Magellanic Cloud
\citep[SMC;][]{Yokogawa2000}, M82 \citep{Griffiths2000} and NGC 3256
(P. Lira et al., in preparation). The LFs of these three galaxies all
appear to follow the same trend such that $N \propto L_{\rm
X}^{-0.65}$. However, the LF of Circinus shows a distinct ``kink''
around $10^{38}-10^{39}$~erg~s$^{-1}$ that is not evident in the other LFs.
\centerline{\includegraphics[width=9cm]{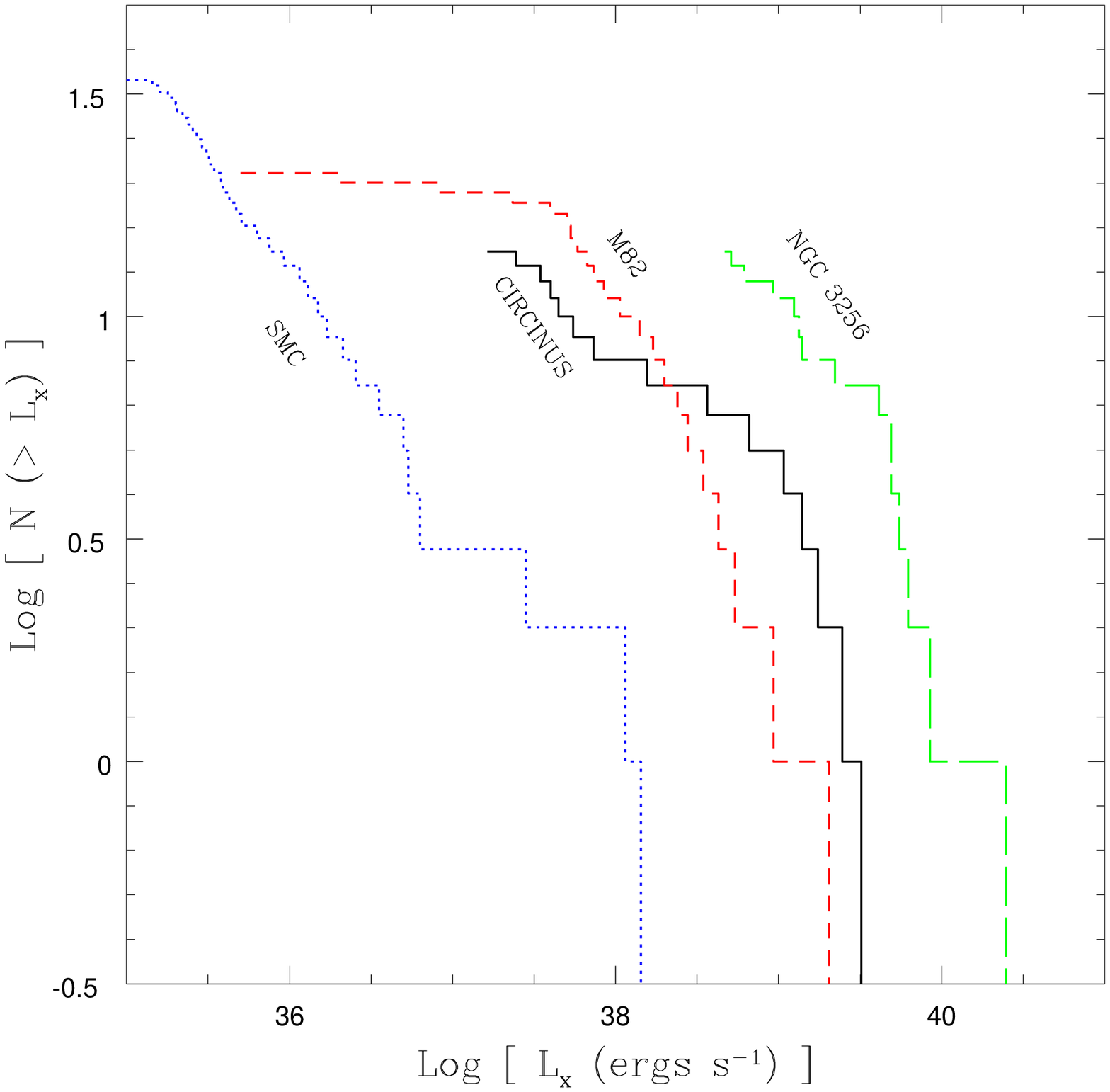}}
\figcaption[fig8.eps]{
Cumulative number of sources versus 0.5-10.0~keV luminosity for
Circinus (solid), the SMC (dotted), M82 (short-dashed), and NGC 3256
(long-dashed). The sources in the SMC, M82 and NGC~3256 all appear to
increase with decreasing luminosity such that $N \propto L_{\rm
X}^{-0.65}$. The Circinus sources do not follow this trend (see
$\S$\ref{lf}).\label{fig:lf}}
\vspace{0.2in}
\noindent One possible explanation for this anomalous LF may be the large
attenuation of soft X-ray photons from both the large Galactic column
and absorption in the HETGS. For example, if the Circinus source
population has a LF slope similar to that of the other galaxies, but
the low-luminosity sources in Circinus are substantially softer than
the high-luminosity ones (an effect seen in NGC~4038/4039; A.~Zezas et
al. 2001, in preparation), a large fraction of the faint, soft X-ray
sources might go undetected. 

To test this hypothesis, we assumed that, in the absence of Galactic
and HETGS absorption, the LF of Circinus has the same average
intrinsic slope as that of the other galaxies in Figure~\ref{fig:lf}
(i.e., $N \propto L_{\rm X}^{-0.65}$). Since the brightest source
detected in Circinus has a 0.5--10.0~keV luminosity of
$3.4\times10^{39}$~erg~s$^{-1}$, we should expect to detect
$\approx$~28 unabsorbed sources at the luminosity limit of our HETGS
observation (i.e., $L_{\rm 0.5-10.0~keV}=2\times
10^{37}$~erg~s$^{-1}$). For simplicity, we assumed that all sources
below $L_{\rm 0.5-10.0~keV} = 1\times 10^{38}$~erg~s$^{-1}$ were
well-fit by a power-law spectrum with an average photon index
$\Gamma$. We then used PIMMS \citep{Mukai2000} to estimate the number
of sources that would fall below our flux threshold as we varied the
average $\Gamma$ from 1.76 (i.e., what we find for the bright sources)
to softer values, assuming an HETGS observation and an average
neutral hydrogen absorption column of $N_{\rm
H}=1.5\times10^{22}$~cm$^{-2}$. We found that in order to reproduce
the observed 15 off-nuclear X-ray sources, an average photon index of
$\Gamma \approx 3.2$ is necessary for the fainter sources. 

While this hypothesis provides a viable explanation for the ``kink''
seen in the LF of Circinus, the average spectral index required for
the fainter sources is significantly softer than the softest spectral
index directly measured via spectral fitting analysis of the brightest
sources in the field (see $\S$~\ref{zero} and Table~\ref{sources}).
This large contrast argues against such a hypothesis, but we note that
a similar trend of systematic softening, although considerably milder,
is seen in {\it Chandra} point-source population studies of nearby
galaxies such as M81 \citep{Tennant2001}, NGC~4038/4039 (A.~Zezas et
al. 2001, in preparation), and NGC~4647 \citep{Sarazin2000}. The small
number of counts in these studies, however, allow only crude hardness
ratio determinations. Such a trend is not found in a deep {\it
XMM-Newton} study of M31 \citep{Shirey2001}, where there are adequate
counts to perform detailed spectral analysis on all sources above
$L_{\rm 0.5-10.0~keV}\approx1\times 10^{37}$~erg~s$^{-1}$.
Unfortunately, a large statistical sample is not yet available to
clarify whether this trend of systematic spectral softening is common
among the fainter point sources in nearby galaxies or whether an
alternate explanation for LF of Circinus exists (e.g., there is some
intrinsic difference between Circinus and the other galaxies such as
morphological type, gas mass, star formation rate, star formation
history, etc. which could explain the ``kink'').

\section{Discussion and Summary}\label{discussion}

We have performed a systematic analysis of 16 point sources
coincident with the disk of Circinus in a {\it Chandra} HETGS
observation, down to a limiting X-ray luminosity of
$2\times10^{37}$~erg~s$^{-1}$. We find that

\begin{itemize}
\item Nearly half of the sources coincide with the star-forming
features of the galaxy, and only four off-nuclear sources have optical
counterparts.

\item Four sources exhibit short-term ($\sim$~1~day) variability, and
one source exhibits only long-term ($\sim$~4~yr) variability.

\item The X-ray spectra of these objects are best-fit by a large
absorption column above the Galactic value and a range of power-law
photon indices from $\Gamma =$ 0.7--2.7.
\end{itemize}

\noindent Based on the large observed absorption columns, spatial
coincidence with the Circinus galaxy, and large X-ray-to-optical flux
ratios, we conclude that the bulk of these off-nuclear sources are XBs
and/or luminous SNRs located within Circinus. The shape of the
Circinus LF appears to be different from those of other galaxies and
may indicate a bias against faint soft X-ray sources introduced by the
large Galactic column and absorption in the HETGS. Any definite
conclusions about the Circinus LF, however, should wait until a
large, complete sample of nearby galaxies can be assembled for
comparision.

The two most luminous off-nuclear sources, CG~X-1 and CG~X-2, are
remarkable. The distinct periodic variability, high X-ray luminosity,
and soft X-ray spectrum of CG~X-1 provide strong support for an
interpretation as a $>50$ $M_{\odot}$ black hole in an accreting
binary system in Circinus. High-luminosity X-ray sources such as this
one are fairly common among nearby galaxies (see $\S$\ref{intro}), but
the physical nature of these objects is often poorly determined
because of the limited photon statistics. The X-ray properties of the
periodic source CG~X-1 provide the strongest evidence to date that at
least some of the super-Eddington X-ray sources detected in other
galaxies are individual XBs with large inferred black-hole masses,
challenging both stellar evolution and accretion disk models
\citep[e.g.,][]{Taniguchi2000, Watarai2001}. Unfortunately, the lack
of an optically detectable stellar companion limits the degree to
which we can constrain the physical parameters of this system. Further
X-ray observations of CG~X-1 are important to tighten constraints on
the variability of its period, luminosity and spectrum. If this source
indeed undergoes periodic eclipses, then the period and shape of the
light curve should remain fairly stable. However, if the apparent
periodicity is due to accretion disk instabilities, one might expect
the X-ray light curve to ``evolve'' away from what we see in
Figure~\ref{fig:periodic}.

Likewise, the large X-ray luminosity, the presence of a very large
equivalent width iron K-shell emission line, the hard X-ray spectrum,
and the association with both a radio point source and an
$\mathrm{H}\alpha$-detected \ion{H}{2} region make CG~X-2 a prime
candidate for a young SNR in Circinus. Continued X-ray monitoring of
this source will determine how its luminosity and spectral properties
evolve with time. Nearby, X-ray-bright supernovae like CG~X-2 are
relatively rare \citep{Schlegel1995, Houck1998}, but they
represent an excellent opportunity to study the physical interaction
of a supernova with a dense circumstellar environment.

The ubiquity of super-Eddington off-nuclear X-ray sources in nearby
galaxies has become an established but poorly understood phenomenon.
Our findings for Circinus support the hypothesis that the off-nuclear, 
super-Eddington X-ray sources in nearby galaxies are the product of
{\it both} young SNRs and intermediate-mass black holes in XBs.

\acknowledgements

We are grateful to D. Alexander, M. Eracleous, S. Gallagher, and R.
Wade for helpful discussions. We thank 
M.~Elmouttie for information regarding radio point sources in Circinus,
R.~Evans and A.~Williams for information on the supernova
monitoring of Circinus,
Y.~Maeda for providing deep Galactic Ridge source counts prior to
publication, 
A.~Marconi and E.~Oliva for kindly sharing their optical images of
Circinus, 
B.~McLean and J.~Garcia for photographic emulsion transmission curves,
and 
B.~Wilkes for assistance with CXC data reprocessing.
Support for this work was provided by
the National Aeronautics and Space Administration through NASA grant
NAS8-38252 (GPG, PI), NASA LTSA grant NAG5-8107 (FEB, WNB, SK) and
{\it Chandra} Grant GO0-1160X (FEB, WNB).

\end{document}